\documentclass[3p,times,procedia]{elsarticle}
\usepackage{nupha_ecrc}

\volume{00}

\lastpage{4}
\firstpage{1}

\journalname{Nuclear Physics A}

\runauth{}

\jid{nupha}

\jnltitlelogo{Nuclear Physics A}

\usepackage{amssymb}
\usepackage{amsthm}
\usepackage{amsmath}
\usepackage{hyperref}
\hypersetup{
	colorlinks=true
}
\usepackage{graphicx}
\usepackage{subcaption}
\newcommand{\trento}{T\raisebox{-0.5ex}{R}ENTo}

\usepackage[figuresright]{rotating}

\newcommand{\be}{\begin{eqnarray}}
\newcommand{\ee}{\end{eqnarray}}
\newcommand{\ma}{\mathrm}
\newcommand{\ml}{\mathcal}
\newcommand{\bs}{\boldsymbol}

\begin{document}

\begin{frontmatter}



\dochead{XXVIIth International Conference on Ultrarelativistic Nucleus-Nucleus Collisions\\ (Quark Matter 2018)}

\title{Quarkonium production in heavy ion collisions: coupled Boltzmann transport equations}


\author{Xiaojun Yao\corref{cor1}}
\cortext[cor1]{xiaojun.yao@duke.edu}
\author{Weiyao Ke}
\author{Yingru Xu}
\author{Steffen Bass}
\author{Berndt M\"uller}
\address{Department of Physics, Duke University, Durham, NC 27708, USA}

\begin{abstract}
We develop a set of coupled Boltzmann equations to describe the dynamical evolution of heavy quarks and quarkonia inside the quark-gluon plasma. The quarkonium dissociation and recombination terms are calculated from pNRQCD. Their interplay drives the system to a detailed balance. The heavy quark energy loss term is necessary for the system to reach kinematic thermalization. By coupling the transport equations with initial particles' momenta generated by \textsc{Pythia} and hydrodynamic medium evolutions, we can describe the $R_{AA}$ of $\Upsilon$ family at both RHIC and LHC energies. The transverse momentum azimuthal anisotropy of $\Upsilon$(1S) in $2.76$ TeV peripheral Pb-Pb collisions is also studied.

\end{abstract}


\end{frontmatter}


\section{Introduction}
\label{sect: intro}

Heavy quarkonium, bound state of heavy quark antiquark pair, can be used as a probe of the quark-gluon plasma (QGP), a hot nuclear environment produced in heavy ion collisions. Due to the Debye (static) screening effect of the plasma, the color attraction between the heavy quark antiquark pair is significantly suppressed at high temperature so that bound state cannot be formed and ``melts" \cite{Matsui:1986dk}. Therefore, it is expected that heavy quarkonium production in heavy ion collisions is suppressed with respect to that in proton-proton collisions. Furthermore, since higher excited quarkonium states are less bound and melt at lower temperatures, quarkonium can be thought of a thermometer for the QGP. However, this simple Debye screening picture fails to describe the experimental measurement of the suppression factor and the extraction of the melting temperature of each quarkonium state from experimental measurements is complicated. Several other crucial effects have to be taken into account: cold nuclear matter (initial state) effects, static and dynamic screening, in-medium dissociation and recombination and feed-down processes. One has to include all these factors in a consistent way.

To this end, we develop a set of coupled Boltzmann transport equations of heavy quarks and quarkonia. It includes elastic and inelastic scattering of heavy quarks with medium particles, color screening, as well as quarkonium dissociation and recombination inside the medium. The dissociation and recombination processes are calculated in potential non-relativistic QCD (pNRQCD). In Sect.~\ref{sect:boltzmann}, the set of coupled Boltzmann equations is introduced and the calculation of dissociation and recombination terms from pNRQCD is briefly explained. Then it follows in Sect.~\ref{sect:equilibrium} some test simulations that demonstrate how the system of heavy quarks and quarkonia approach detailed balance and equilibrium. Comparisons with experimentally measured nuclear modification factors $R_{AA}$ are shown in Sect.~\ref{sect:results}. Finally, a short conclusion is drawn in Sect.~\ref{sect:conclusion}.

\section{Coupled Boltzmann Equations}
\label{sect:boltzmann}
The set of coupled Boltzmann equations is given by \cite{Yao:2017fuc}
\be
\label{eq:LBE}
(\frac{\partial}{\partial t} + \dot{{\bs x}}\cdot \nabla_{\bs x})f_Q({\bs x}, {\bs p}, t) &=& \ml{C}_Q  -  \ml{C}_Q^{+} +  \ml{C}_Q^{-}\\
(\frac{\partial}{\partial t} + \dot{{\bs x}}\cdot \nabla_{\bs x})f_{\bar{Q}}({\bs x}, {\bs p}, t) &=& \ml{C}_{\bar{Q}}  -  \ml{C}_{\bar{Q}}^{+} + \ml{C}_{\bar{Q}}^{-}\\
(\frac{\partial}{\partial t} + \dot{{\bs x}}\cdot \nabla_{\bs x})f_{nl}({\bs x}, {\bs p}, t) &=& \ml{C}_{nl}^{+}-\ml{C}_{nl}^{-} \,,
\ee
where $f_i({\bs x}, {\bs p}, t)$ denotes the phase space distribution function for the heavy quark $i=Q$, antiquark $i=\bar{Q}$ and each quarkonium state with quantum number $nl$. Polarizations of quarkonia are not considered here. The left-hand side of the equations describes the free streaming of particles while the right-hand side contains interactions between heavy particles and light quarks and gluons (abbreviated as $q$ and $g$) in the hot medium. For heavy quarks $Q$, the collision term $\ml{C}_Q$ includes three types of scattering processes: the elastic ${2\rightarrow2}$, inelastic ${2\rightarrow3}$ and ${3\rightarrow2}$ scatterings. In this work, we use a specific perturbative calculation and numerical implementation in Ref.~\cite{Ke:2018tsh}.

\begin{figure}[h]
    \centering
    \begin{subfigure}[t]{0.33\textwidth}
        \centering
        \includegraphics[height=1.0in]{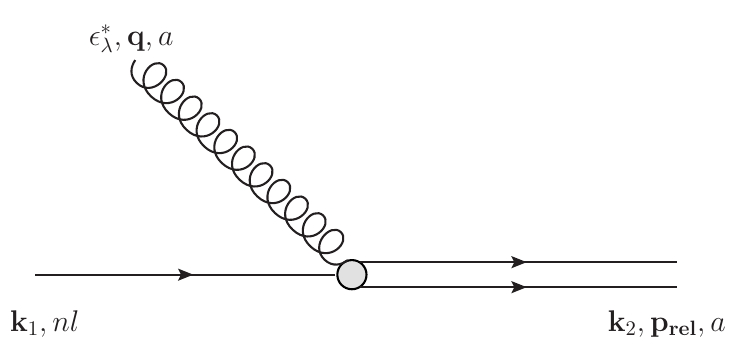}
        \caption{gluon absorption/emission}
    \end{subfigure}%
    ~ 
    \begin{subfigure}[t]{0.33\textwidth}
        \centering
        \includegraphics[height=1.3in]{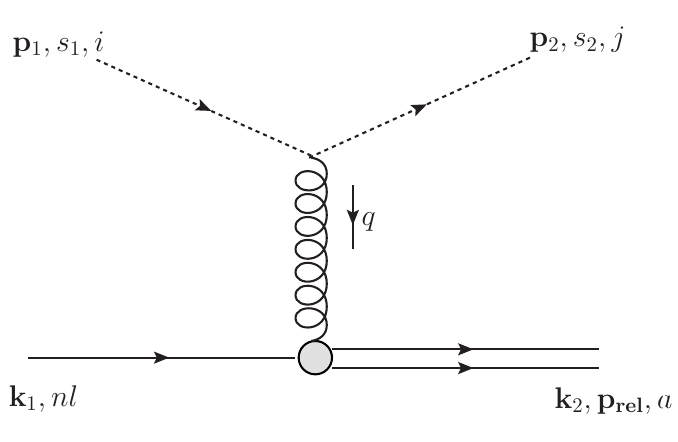}
        \caption{inelastic scattering with light quarks}
    \end{subfigure}%
    ~ 
    \begin{subfigure}[t]{0.33\textwidth}
        \centering
        \includegraphics[height=1.3in]{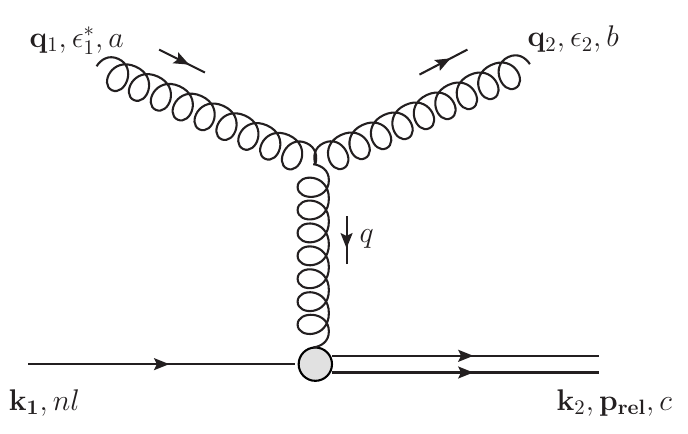}
        \caption{inelastic scattering with gluons}
    \end{subfigure}
    \caption{Tree-level Feynman diagrams contributing to dissociation and recombination up to NLO in coupling constant. Single solid line denotes quarkonium while double solid lines represent the unbound heavy quark pair.}
    \label{fig:scattering}
\end{figure}

The quarkonium dissociation and recombination terms $C_i^{\pm}$ are calculated from pNRQCD, which relies on systematic nonrelativisitc and multipole expansions. The pNRQCD has been used to study quarkonium dissociation rate \cite{Brambilla:1999xf, Brambilla:2011sg, Brambilla:2013dpa}. Here we assume $M\gg Mv \gg Mv^2, T, m_D$ so that bound state can be formed. Here $M$ is the heavy quark mass, $v$ is the relative velocity of the heavy quarks inside the bound state, $T$ denotes the plasma temperature and $m_D$ is the Debye mass. Each quarkonium state has a specific melting temperature above which recombination cannot occur. Relevant tree-level diagrams up to next-to-leading (NLO) order in the coupling constant are calculated, as shown in Fig.~\ref{fig:scattering}. Explicit expressions of the scattering amplitudes and collision terms in the Boltzmann equations can be found in Ref.~\cite{xy_bm}.

\section{Detailed Balance and Equilibrium}
\label{sect:equilibrium}
We solve the Boltzmann equations by Monte Carlo simulations. From now on, we will focus on the bottom quarks and bottomonia because the $b$ quark mass is large and the expansions of the effective field theory are justified. First we conduct test simulations where the medium is given by a QGP box. The QGP box has side length $10$ fm and the plasma temperature is constant throughout the box. Fixed numbers of $b$, $\bar{b}$ and $\Upsilon$(1S) are sampled as the initial condition. Their positions are sampled randomly inside the QGP box. Their momenta can be sampled from either thermal distributions or uniform distributions. Periodic boundary conditions are applied to the system. 

\begin{figure}[h]
    \centering
    \begin{subfigure}[t]{0.5\textwidth}
        \centering
        \includegraphics[height=1.8in]{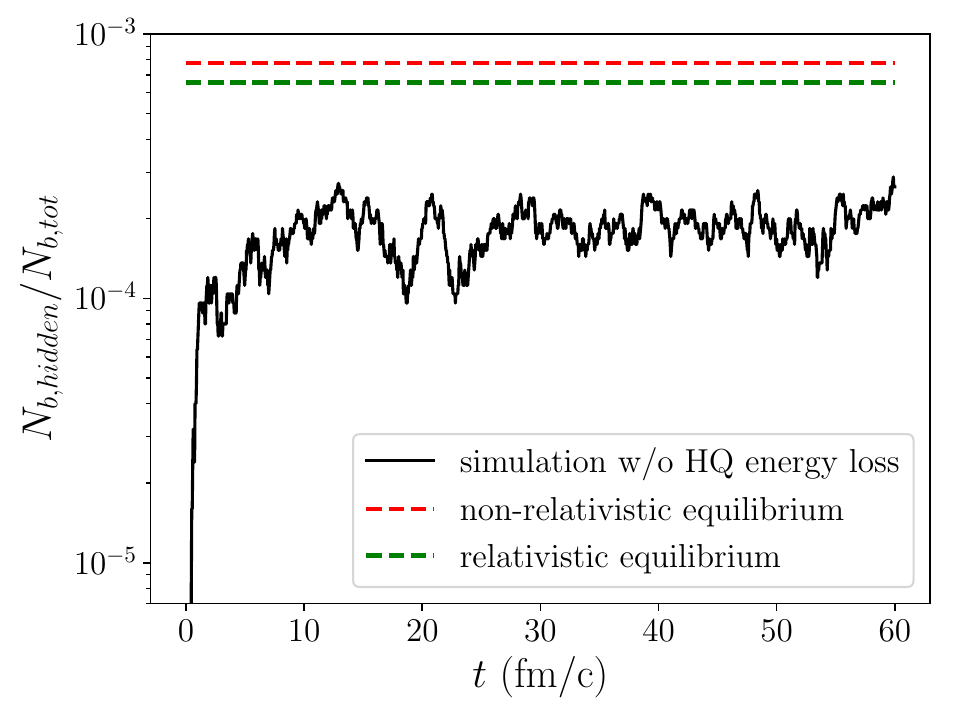}
        \caption{Simulation with heavy quark energy loss turned off.}
    \end{subfigure}%
    ~ 
    \begin{subfigure}[t]{0.5\textwidth}
        \centering
        \includegraphics[height=1.8in]{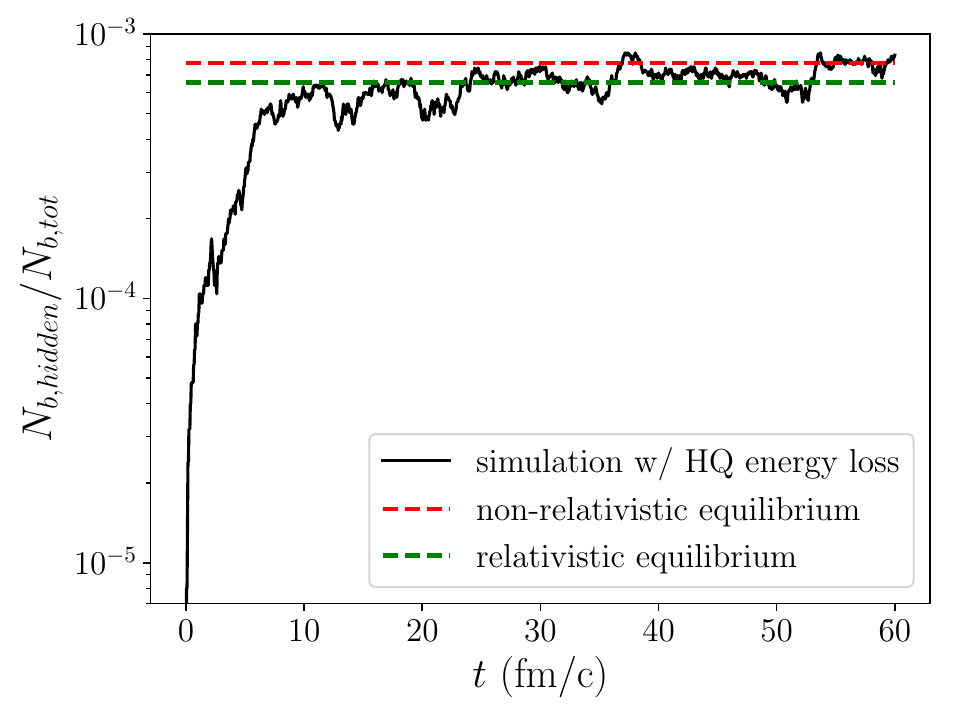}
        \caption{Simulation with heavy quark energy loss turned on.}
    \end{subfigure}
    \caption{Hidden $b$-flavor fractions from simulations at $T=300$ MeV with $N_{b,\ma{tot}} =25$ inside a QGP box.}
    \label{fig:balance}
\end{figure}

Simulation results at $T=300$ MeV with $N_{b,\ma{tot}} = N_b+N_{\Upsilon(1S)} = 25$ are shown in Fig.~\ref{fig:balance}.
The $p_x,p_y,p_z$ of each $b$ quark are sampled from a uniform distribution between $0$ and $5$ GeV. Two situations are studied here: one with the heavy quark energy loss term turned off and the other with that turned on.
The fraction of hidden $b$-flavor ($\frac{N_{b,\ma{hidden}}}{N_{b,\ma{tot}} }$ = $\frac{N_{\Upsilon(1S)}}{N_{b,\ma{tot}} }$) is plotted as a function of time and compared with that at equilibrium. The hidden $b$-flavor fraction at equilibrium can be calculated as in Ref.~\cite{Yao:2017fuc}. Here the degeneracy between $\Upsilon$(1S) and $\eta_b$ is lifted up in the recombination.

In both situations, the hidden $b$-flavor fraction approaches a static value. The interplay between dissociation and recombination drives the system to a detailed balance. Furthermore, without the heavy quark energy loss, the system reaches a state differing from equilibrium while with the energy loss included, it reaches the proper equilibrium state. Therefore the heavy quark energy loss term is necessary to drive the system to a kinematic thermalization, since no term analogous to the heavy quark energy loss is included in the quarkonium Boltzmann equation.

\section{Results}
\label{sect:results}
\begin{figure}[h]
    \centering
        \begin{subfigure}[t]{0.33\textwidth}
        \centering
        \includegraphics[height=1.5in]{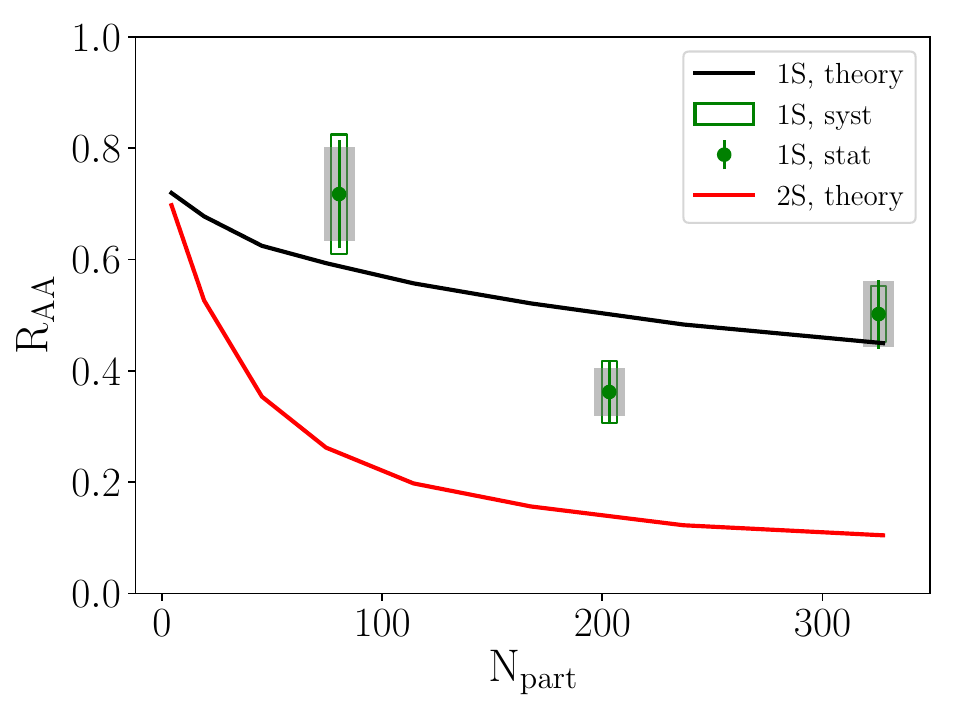}
        \caption{$R_{AA}$ as a function of centrality and STAR measurements.}
    \end{subfigure}%
    ~ 
    \begin{subfigure}[t]{0.33\textwidth}
        \centering
        \includegraphics[height=1.5in]{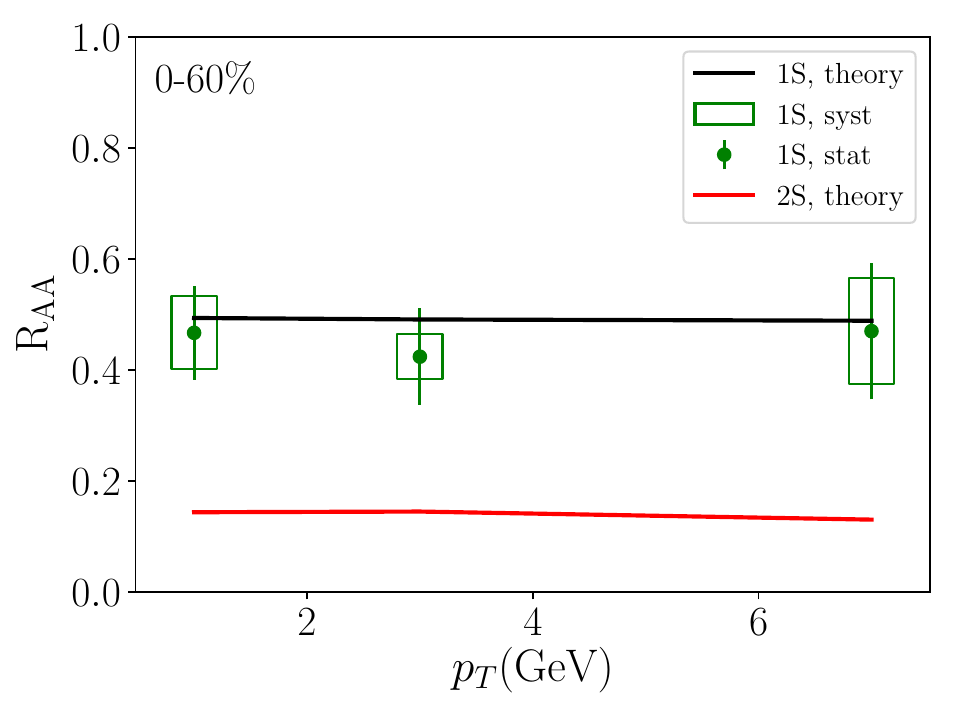}
        \caption{$R_{AA}$ as a function of transverse momentum and STAR measurements.}
    \end{subfigure}%
    ~
    \begin{subfigure}[t]{0.33\textwidth}
        \centering
        \includegraphics[height=1.5in]{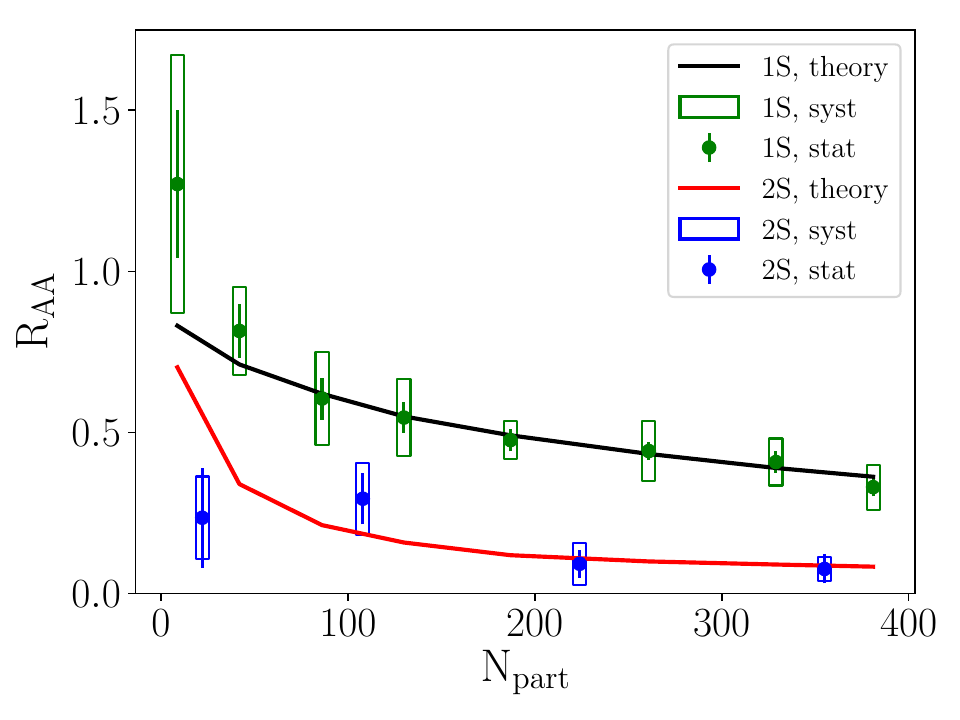}
        \caption{$R_{AA}$ as a function of centrality and CMS measurements.}
    \end{subfigure}
    ~ 
    \begin{subfigure}[t]{0.33\textwidth}
        \centering
        \includegraphics[height=1.5in]{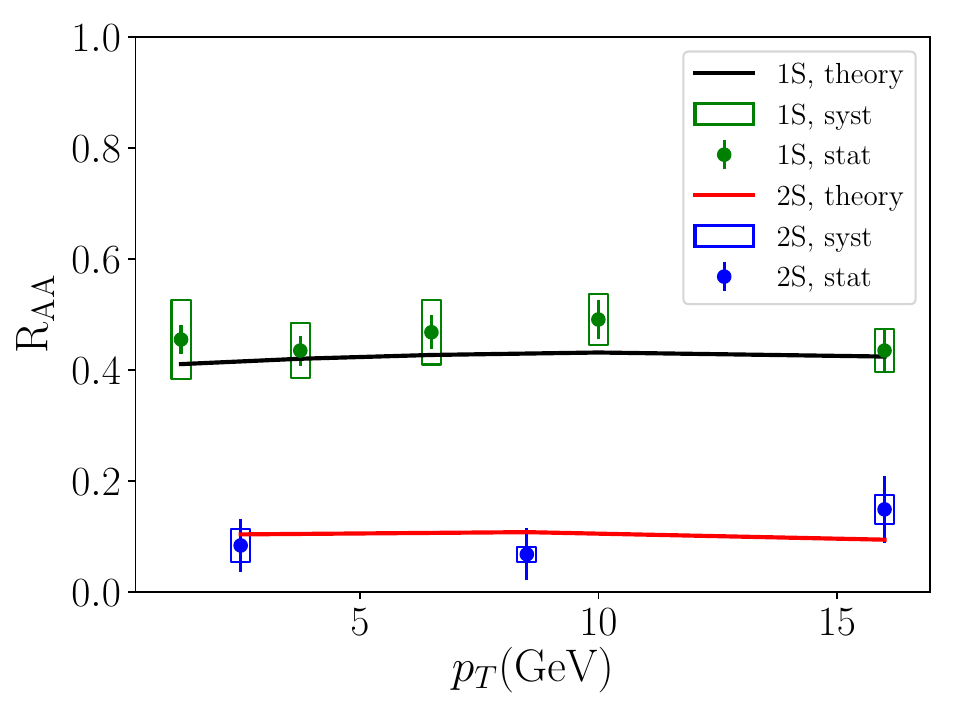}
        \caption{$R_{AA}$ as a function of transverse momentum and CMS measurements.}
    \end{subfigure}%
    ~
        \begin{subfigure}[t]{0.33\textwidth}
        \centering
        \includegraphics[height=1.5in]{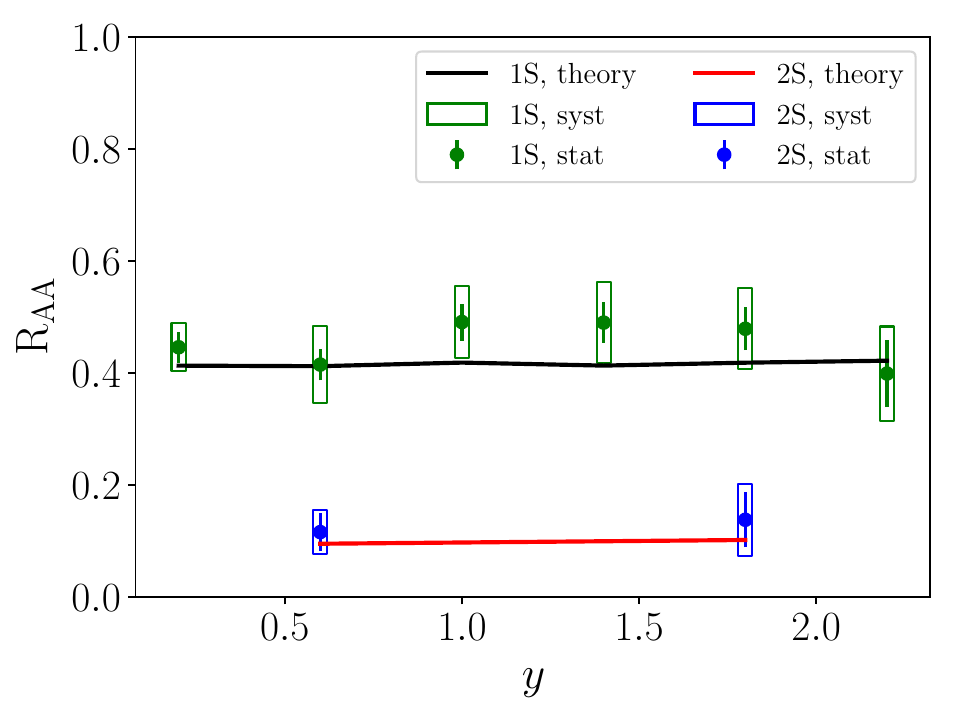}
        \caption{$R_{AA}$ as a function of rapidity and CMS measurements.}
    \end{subfigure}%
    ~ 
    \begin{subfigure}[t]{0.33\textwidth}
        \centering
        \includegraphics[height=1.5in]{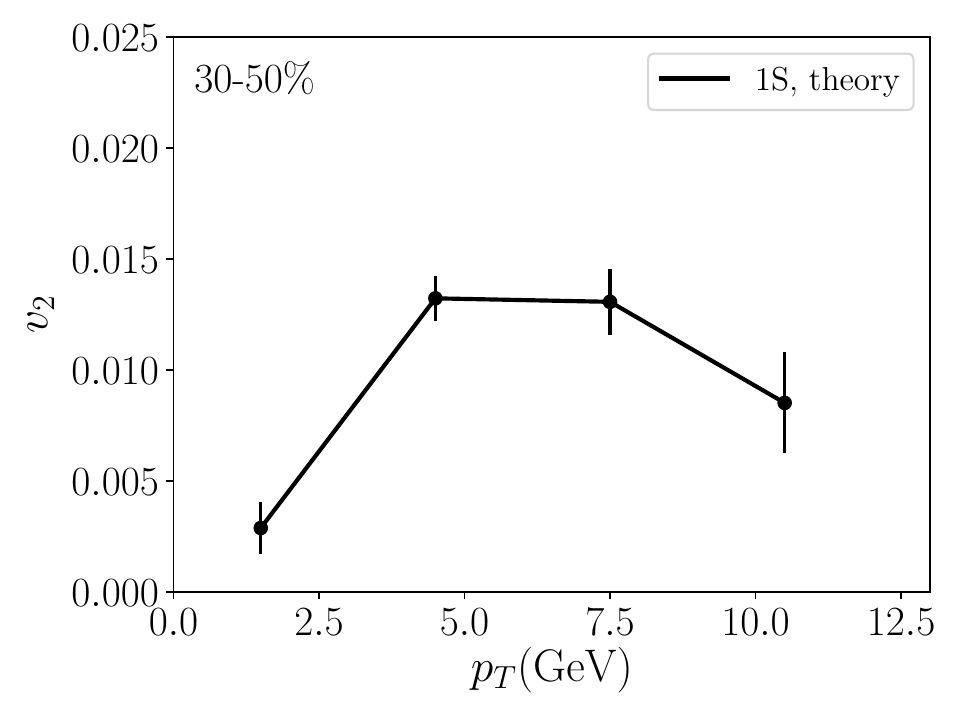}
        \caption{$v_2$ of $\Upsilon$(1S) in $2.76$ TeV Pb-Pb collisions with centrality 30-50\%.}
    \end{subfigure}%
    
    \caption{$R_{AA}$ and $v_2$ in $200$ GeV Au-Au and $2.76$ TeV Pb-Pb collisions. Data are taken from Ref.~\cite{Ye:2017fwv, Khachatryan:2016xxp}.}
    \label{fig:star_cms}
\end{figure}

We study the $\Upsilon$ production in $200$ GeV Au-Au and $2.76$ TeV Pb-Pb collisions in mid-rapidity. We include $\Upsilon$(1S) and $\Upsilon$(2S) in the Boltzmann equations. The melting temperature of  $\Upsilon$(2S) is set to be $210$ MeV above which no $\Upsilon$(2S) state can be formed. The coupling constant is fixed to be $\alpha_s=0.3$ and the attractive potential is $V_S=-\frac{0.56}{r}$. The initial momenta of particles generated from hard scattering are sampled from the event generator \textsc{Pythia} \cite{Sjostrand:2014zea} with nuclear parton distribution function \cite{Eskola:2009uj}. The positions of initial hard scattering vertices are sampled according to the binary collision density calculated in \trento. {\trento} also generates the initial entropy density for the 2+1D relativistic viscous hydrodynamic evolution \cite{Shen:2014vra}. The initial condition and medium properties are calibrated to the soft hadron observables \cite{Bernhard:2016tnd}. Event-averaged hydrodynamics is used. The cold nuclear matter suppression factor is assumed to be $0.72$ and $0.87$ for $200$ GeV Au-Au and $2.76$ TeV Pb-Pb collisions separately, which modifies the overall production rate. The branching ratio of $\Upsilon$(2S) to $\Upsilon$(1S) in hadronic phase is $0.26$. The calculated $R_{AA}$ is consistent with experimental data, as shown in Fig.~\ref{fig:star_cms}. Including other excited bottomonium states will on one hand further suppress $\Upsilon$(1S) production because of less feed-down; On the other hand, the dissociated excited state may recombine as $\Upsilon$(1S) state and thus enhancing its production.
The $\Upsilon$(1S) transverse momentum azimuthal anisotropy in $2.76$ TeV peripheral Pb-Pb collisions is also studied and an estimate of $v_2$ is given in Fig.~\ref{fig:star_cms}. We will learn more about in-medium evolutions of bottom quarks and bottomonia from $v_2$ measurements.

\section{Conclusion}
\label{sect:conclusion}
We use a set of coupled Boltzmann equations to describe the in-medium evolution of heavy quarks and quarkonia. By choosing appropriate parameter values, we can explain the experimentally measured $R_{AA}$ of $\Upsilon$ family as functions of centrality, rapidity and transverse momentum. An estimate of the $\Upsilon$(1S) $v_2$ is also presented. The same framework can also be generalized to study the production of doubly charm baryon \cite{Yao:2018zze}. In future, we will use temperature-dependent potentials motivated from lattice calculations, include other quarkonium states (1P, 2P, 3S) in the Boltzmann equations and use event-by-event hydrodynamic simulations to give a more complete understanding of quarkonium production in heavy ion collisions.

This work is supported by U.S. Department of Energy under Research Grant No. DE-FG02-05ER41367. X.Y. also acknowledges support from Brookhaven National Laboratory.




\bibliographystyle{elsarticle-num}



\end{document}